\documentclass{article}
\usepackage[T1]{fontenc}
\usepackage[latin9]{inputenc}
\usepackage[english]{babel}
\usepackage{fancybox}
\usepackage{calc}
\usepackage{textcomp}
\usepackage{amssymb}
\usepackage[pdftex]{graphicx}
\usepackage{setspace}
\usepackage[authoryear]{natbib}
\PassOptionsToPackage{normalem}{ulem}
\usepackage{ulem}
\usepackage{subscript}
\usepackage[unicode=true,pdfusetitle,
 bookmarks=true,bookmarksnumbered=false,bookmarksopen=false,
 breaklinks=false,pdfborder={0 0 1},backref=false,colorlinks=false]
 {hyperref}

\makeatletter

\newcommand{\noun}[1]{\textsc{#1}}
\providecommand{\tabularnewline}{\\}

\newcommand{\lyxaddress}[1]{
	\par {\raggedright #1
	\vspace{1.4em}
	\noindent\par}
}

\makeatother

\begin{document}
\title{Possibility conditions for Open Access}
\author{Jacinto D{\'a}vila}
\maketitle

\lyxaddress{\textsuperscript{1}Centro de Simulaci{\'o}n y Modelos, CESIMO, de la
Universidad de Los Andes y Comunidad de Aprendizaje en Servicios P{\'u}blicos
de la Universidad Polit{\'e}cnica Territorial de M{\'e}rida. Venezuela}
\begin{abstract}
This is an attempt to formalize the conditions of possibility for
free, libre, open access to scientific knowledge within a game. The
challenge is to enunciate the terms under which agents participating
in the \textquotedbl Grand conversation\textquotedbl{} of science
would be willing to open share, exchange, negotiate or surrender their
contributions, considering their correspon\-ding intentions,
goals, beliefs and expected utilities. Many conclusions can be drawn
from the game here described. We have made many simplifying decisions
along the modelling process that must be taken into account as a determining
context for those conclusions, of course. It can be safely state,
however, that under the current conditions of the game, Editors will
keep betting on Toll Access, knowledge distribution models even if
all the other Academic agent go for Open Access.
\end{abstract}

\section{Introduction}

This is an attempt to formalize the conditions of possibility for
free, libre, open access to scientific knowledge within a game. The
challenge is to enunciate the terms under which agents participating
in the \textquotedbl Grand conversation\textquotedbl{} of science
would be willing to open share, exchange, negotiate or surrender their
 contributions, considering their corresponding intentions,
goals, beliefs and expec\-ted utilities. 

We have been studying the possibility, suitability and possible success
of modeling the forms of advocacy of, and resistance to, Open Access
to the scientific literature according to game theory methodology.
While this approach was deemed to be both highly innovative and original,
the simplifications need to make this approach work were judged to
be too extensive to design a realistic model with the available information.
As an alternative, it was decided that doing a thorough survey of
points and terms of resistance as they appear according to the stakeholders,
namely researchers, graduate students, junior faculty, senior faculty,
administrators, funders of research, publishers, libraries and politicians,
would form a useful first step toward the original objective. Also,
it has been deemed crucial to distinguish between developing, emergent
and developed countries, hoping to address complexities and accounting
for local idiosincracies in the process. Having done so, however,
we were faced with the enormous complexity of the global situation
and were forced to make the simplifications that are used and discussed
in this paper. 

The paper is organiced as follows: Firstly, an update on the history
of science and technology, particularly with regard to scientific
publishing. In section 3 we summarize the development of a definition
for open access. Section 4 develops the view of open access in science
as a game of interests. The last section presents the conclusions. 

\section{An update on the history of science and techno\-logy, particularly
with regard to scientific pu\-blish\-ing. }

Publishing is a crucial action for the existing setup of global science.
In a very simplified view, it correspond to the stage in which the
results of research and experimentation are communicate to everybody
with the primary objective of enlightening people and, therefore,
helping to solve their problems. Seeing science as a open and global
enterprise entails, however, a secondary objective for publishing:
the results must be verified and validated by other scientists all
over the world, so as to establish the quality and general validity
of those results. And a third objective in this sequence is to allow
those results to be made available for further research, to empower
those that are willing to keep researching the corresponding family
of problems. 

This simplified view of the processes of science related to publishing,
which is almost all that is taught about the history of science to
most new scientist, is actually too simple in many ways. In the most
established of western views, for once, science do not simply progress
by validation of general results, but it also advanced by refutation
of hypothesis aspiring to be general results\citet{lakatos1976}.
The urge to check on somebody else's results is not (only) to do the
good deed of validating a work that can be trusted, but
an interest in finding lacks or cracks in their proposed explanation.
That is to refute them and establish their conclusion (and their model)
as invalid and, therefore, to force the search for a new model. This
has the further implication of establishing that positive and negative
results are both very important for the advancement of science. All
these interactions over positive and negative results create the layout
for the grand conversation of science which must be published for it
own sake. 

But there is another dimension in which that view is too simple. It
is the implicit assumption that there are equally meaningful families
of problems to which scientists suscribe and try to solve as a team
for the benefit of everybody. It is equivalent to the very na�ve assumption
that there exist a supreme, neutral goal for science and that scientist
should and will be equally motivated to work in their chosen family
of problems disregarding the local circunstances in which their (the
scientists and their poblems) are embedded. 

The truth is that science is heavily influenced by technological,
economics and political dinamics. In the western societies, for instance,
it is normally expected that a scientific development will turn into
a technological solution and, eventually, a product or a service that
the public can consume to solve their needs. And which needs must
be solved first is, of course, a collective choice. A market choice
for some, but in any case, a political decision (which is not the
same as a decision made by politicians, even though it can be). 

Scientific publishing, in particular, has been influenced by economical
con\-cerns since very early in its history. As explained in \citet{guedon2001},
the very origins of the concept of \textbf{intelectual property} can
be traced back to a (very succesful) attempt to extend the concept
of landed property: ``This bit of legal creativity was actually motivated
by the stationers who needed to establish legally viable claims over
the texts they printed, if only to protect their trade from imitation
and piracy. To them, this meant exclusive and perpetual owner\-ship,
as is the case for land property. But \textbf{they were not the only
players} and, as a result of various court actions, the definition
of what they actually claimed to own remained murky for several decades,
almost a century, actually'' (.ibid, chapter 3, our emphasis). 

We will not say more about the history of printed documents except
that, with the arrival of Internet and the XXI century electronic
technology, making a copy of a book became a process of almost cero
cost (apart from the costs of producing the book in the first place).
In this context, the actual value of the text (images included) can
be almost completely mapped to its content and, eventually, to its
authors. This new situation has apparently triggers a wave of questions
about the origins of that value and of who are entitled to use and
enjoy each contribution. A line of questioning that is particulary
striking for the scientific practice, where scientists constantly
look for accessing ideas and proposals by other scientists while trying
to solve scientific problems. 

\section{The development of a definition for open access. }

Peter Suber in \citet{PS2012} explains the effects of that shift
from ink on paper into digital or electronic texts. This book relates
that ideal conditions for the grand conversation of science: authors
that are free to read and share their ideas in the process of designing,
establishing, reporting and refuting theories and experiments. Those
conditions we call Open Access, OA, and they can be summarized like
this:

``What is open access? Open access means that scientific literature
should be publicly available, free of charge on the Internet so that
those who are interested can read, download, copy, distribute, print,
search, refer to and, in any other conceivable legal way, use full
texts without encountering any financial, legal or technical barriers
other than those associated with Internet access itself''\footnote{\href{http://openaccess.mpg.de/2365/en}{http://openaccess.mpg.de/2365/en}}.

What, in turn, makes OA possible? Suber kindly sumarizes an answer
like this: ``OA is made possible by the internet and copyright-holder
consent. (.ibid, pg 9)''. He goes on to make clear what OA is not
intented for: ``OA isn\textquoteright t an attempt to bypass peer
review {[}..{]}. OA isn\textquoteright t an attempt to reform, violate,
or abolish copyright {[}..{]}. OA isn\textquoteright t an attempt
to deprive royalty-earning authors of income {[}..{]}. OA isn\textquoteright t
an attempt to deny the reality of costs {[}..{]}. OA isn\textquoteright t
an attempt to reduce authors\textquoteright{} rights over their work
{[}..{]}. OA isn\textquoteright t an attempt to reduce academic freedom
{[}..{]}. OA isn\textquoteright t an attempt to relax rules against
plagiarism {[}..{]}. OA isn\textquoteright t an attempt to punish
or undermine con- ventional publishers {[}..{]}. OA doesn\textquoteright t
require boycotting any kind of literature or publisher {[}..{]}. OA
isn\textquoteright t primarily about bringing access to lay readers
{[}..{]}. Finally, OA isn\textquoteright t universal access {[}..{]}''
(.ibid, pg 20--27).

From this brief account it must be clear that copyright-holders' interests
are crucial. This is the reason why we have chosen to study the problem
as a game of interests. 

\section{Open Access in science as a game of interests. Conditions of possibility
for equilibrium. }

Let us formally define the game we are analizing. From mathematical
game theory \citet{leytonbrownshoham2008}, a game is a tuple (N,A,u)
where: 
\begin{itemize}
\item N is a finite set of n players, indexed by i;
\item A = $A_{1}\otimes\cdots\otimes A_{n}$, where A\textsubscript{i}
is a finite set of actions available to player i. Each vector a =
(a\textsubscript{1}, \ldots, a\textsubscript{n}) in A is called
an action profile; 
\item u = (u\textsubscript{1},, \ldots,u\textsubscript{n}) where u\textsubscript{i}:
A \textrightarrow{} \ensuremath{\mathbb{R}} is a real-valued utility
(or payoff) function for player i.
\end{itemize}
The amount of players and the number of possible actions for each
one is important for combinatorial reasons. Thus, let us start by
listing the actual agents involved in the target problem and see if
that that list and the lists of their actions can be simplified. Agents
in the game of scientific knowledge include, at least, the following:
\begin{enumerate}
\item Researchers
\item Graduate students
\item Junior faculty
\item Senior faculty
\item Librarians 
\item Administrators
\item Funders of research
\item Editors
\item Politicians
\end{enumerate}
These are, of course, groups of people distributed all over the world
and with many other particular characteristics. Thus, this is already
a reduction of complexity. We believe, however, that for the problem
of scientific knowledge generation and distribution, those categories
would suffice. But they are still too many when one considers the
number of possible actions for each agent group. So, let us try another
symplification together with the listing of those action possibilities:

\begin{center}
\begin{tabular}{|c|c|}
\hline 
Agents & Actions\tabularnewline
\hline 
\hline 
Academics & %
\begin{tabular}{c}
Publish TA\tabularnewline
Publish OA\tabularnewline
Perish\tabularnewline
\end{tabular}\tabularnewline
\hline 
Administrators & %
\begin{tabular}{c}
Support TA\tabularnewline
Support OA\tabularnewline
Support Both\tabularnewline
\end{tabular}\tabularnewline
\hline 
Funders & %
\begin{tabular}{c}
Demand publications\tabularnewline
Demand OA publications\tabularnewline
Don't demand anything\tabularnewline
\end{tabular}\tabularnewline
\hline 
Editors & %
\begin{tabular}{c}
Grant TA\tabularnewline
Grant OA\tabularnewline
Grant big deals\tabularnewline
Grant OA with embargoes\tabularnewline
\end{tabular}\tabularnewline
\hline 
Politicians & %
\begin{tabular}{c}
Permit TA\tabularnewline
Demand green OA\tabularnewline
Demand gold OA\tabularnewline
Demand some OA\tabularnewline
\end{tabular}\tabularnewline
\hline 
\end{tabular}
\par\end{center}

\begin{center}
Table 1: Players and their actions, second approximation
\par\end{center}

In Table 1, we reduce the list of agents to 5 and present some actions
for each agent type. With this proposal, we are still dealing with
$3^{3}*4^{2}=432$ possibilities. But at this point we can discuss
some general features of this game. We are simplyfing the actual game
in another aspect. By concentrating on actions related to the issue
of access to scientific communications, papers in particular, we manage
to reduce the number of possible actions, but also, let us called
it, the timeframe of the game. It is like an instantaneous game. Although
actions can take extended and different times to be executed, we will
be focused on the net effect of an agent making a move at time that
coincides or overlaps other moves by other agents in the game. This
is, of course, another simplification.

Let us insist that, in this game, each agent represents a group of
people. This is particularly important because it allows us to take
modelling advantage of a normally obscured concept in game theory:
the concept of \textbf{mixed strategy}. A mixed strategy is a linear
combination of pure strategies. And a pure strategy correspond to
a choice of one particular action by the agent. 

For example, ``publish OA'' correspond to the actions of academics
publishing their research as open access documents. It is also, as
such, a pure strategy for the agent Academics in table 1. What would
it be a linear combination of pure strategies?. Some weighted decisions
about those actions. For example, one could say that ``Academics''
selects ``publish TA'' with a probabilty of 0.8 (80\%), ``publish
OA'' with 0.2 (20\%) and never selects ``perish''. In simple agents
game theory, these probabilities are hard to imagine (who is her/his
right mind would play a lotery to make an important decision), but
in our context of groups as agents one could easily attribute those
probabilities to some polling over the members of that game. In the
last example, this would mean that 80 percent of the members of the
group ``Academics'' publish their research as so-called toll access,
20\% as open access and nobody refrain from publish (as this would
be suicidal). 

We were tempted to further simplifying with the assumption that funders
and politicians are one and the same group of people. We have to refrain
from this after noticing that important effects of the actions, also
known as outcomes in game theory, would be underepresented. In particular,
there is a important reduction in the model which have to be balanced:
there is no explicit representation of the whole society, an allegedly
important component of the academic ecosystem as the final receptor
or consumer or, at least, user of knowledge generated by the other
components (a reduction itself, as knowledge could come from other
sources). We decided to deal with this by keeping the politician agent
and modelling the expected outcomes as discrete fields of selected
variables, as shown in table 2:

\begin{center}
{\scriptsize{}}%
\begin{tabular}{|c|c|c|}
\hline 
{\scriptsize{}Agents} & {\scriptsize{}Actions} & {\scriptsize{}Outcomes}\tabularnewline
\hline 
\hline 
{\scriptsize{}Academics} & {\scriptsize{}}%
\begin{tabular}{c}
{\scriptsize{}Publish TA}\tabularnewline
{\scriptsize{}Publish OA}\tabularnewline
{\scriptsize{}Perish}\tabularnewline
\end{tabular} & {\scriptsize{}}%
\begin{tabular}{cc}
{\scriptsize{}Opportunity} & {\scriptsize{}\{}\noun{\scriptsize{}Maximal, Minimal}{\scriptsize{}\}}\tabularnewline
{\scriptsize{}Visibility} & {\scriptsize{}\{}\noun{\scriptsize{}More, Less}{\scriptsize{}\}}\tabularnewline
{\scriptsize{}Prestige} & {\scriptsize{}\{}\noun{\scriptsize{}More, Less}{\scriptsize{}\}}\tabularnewline
{\scriptsize{}Promotion} & {\scriptsize{}\{}\noun{\scriptsize{}More, Less}{\scriptsize{}\}}\tabularnewline
\end{tabular}\tabularnewline
\hline 
{\scriptsize{}Administrators} & {\scriptsize{}}%
\begin{tabular}{c}
{\scriptsize{}Support TA}\tabularnewline
{\scriptsize{}Support OA}\tabularnewline
{\scriptsize{}Support Both}\tabularnewline
\end{tabular} & {\scriptsize{}}%
\begin{tabular}{cc}
{\scriptsize{}Savings} & {\scriptsize{}\{}\noun{\scriptsize{}More, Less}{\scriptsize{}\}}\tabularnewline
\end{tabular}\tabularnewline
\hline 
{\scriptsize{}Funders} & {\scriptsize{}}%
\begin{tabular}{c}
{\scriptsize{}Demand publications}\tabularnewline
{\scriptsize{}Demand OA publications}\tabularnewline
{\scriptsize{}Don't demand anything}\tabularnewline
\end{tabular} & {\scriptsize{}}%
\begin{tabular}{cc}
{\scriptsize{}Quality results} & {\scriptsize{}\{}\noun{\scriptsize{}More, Less}{\scriptsize{}\}}\tabularnewline
\end{tabular}\tabularnewline
\hline 
{\scriptsize{}Editors} & {\scriptsize{}}%
\begin{tabular}{c}
{\scriptsize{}Grant TA}\tabularnewline
{\scriptsize{}Grant OA}\tabularnewline
{\scriptsize{}Grant big deals}\tabularnewline
{\scriptsize{}Grant OA with embargoes}\tabularnewline
\end{tabular} & {\scriptsize{}}%
\begin{tabular}{cc}
{\scriptsize{}Income} & {\scriptsize{}\{}\noun{\scriptsize{}More, Less}{\scriptsize{}\}}\tabularnewline
\end{tabular}\tabularnewline
\hline 
{\scriptsize{}Politicians} & {\scriptsize{}}%
\begin{tabular}{c}
{\scriptsize{}Permit TA}\tabularnewline
{\scriptsize{}Demand green OA}\tabularnewline
{\scriptsize{}Demand gold OA}\tabularnewline
{\scriptsize{}Demand some OA}\tabularnewline
\end{tabular} & {\scriptsize{}}%
\begin{tabular}{cc}
{\scriptsize{}Societal impact \& relevance} & {\scriptsize{}\{}\noun{\scriptsize{}More, Less}{\scriptsize{}\}}\tabularnewline
\end{tabular}\tabularnewline
\hline 
\end{tabular}{\scriptsize\par}
\par\end{center}

\begin{center}
Table 2: Players, actions and outcomes (with the sets of possible
values)
\par\end{center}

Table 2 describes outcomes with a set of variables that can be associated
to the actions of each agent-group. We have chosen the variables that
we believe involved in determining the utility for each agent but,
instead of consolidating a mathematical expression of it, we assign
a finite set of possible values to each variable and, thus, define
a discrete universe of possibilities to explore. As a reference, let
us indicate a couple of possibilities to describe the current situation
and an ideal situation, from the point of view of Open Access. Let
us depict those along with the explanation of the variables and assigned
values. 

\begin{center}
{\footnotesize{}}%
\begin{tabular}{|c|c|c|}
\hline 
{\footnotesize{}Agents} & {\footnotesize{}Actions} & {\footnotesize{}Outcomes}\tabularnewline
\hline 
\hline 
{\footnotesize{}Academics} & {\footnotesize{}}%
\begin{tabular}{c}
{\footnotesize{}Publish TA}\tabularnewline
\end{tabular} & {\footnotesize{}}%
\begin{tabular}{cc}
{\footnotesize{}Opportunity} & \noun{\footnotesize{}Minimal}\tabularnewline
{\footnotesize{}Visibility} & \noun{\footnotesize{}Less}\tabularnewline
{\footnotesize{}Prestige} & \noun{\footnotesize{}More}\tabularnewline
{\footnotesize{}Promotion} & \noun{\footnotesize{}More}\tabularnewline
\end{tabular}\tabularnewline
\hline 
{\footnotesize{}Administrators} & {\footnotesize{}}%
\begin{tabular}{c}
{\footnotesize{}Support TA}\tabularnewline
\end{tabular} & {\footnotesize{}}%
\begin{tabular}{cc}
{\footnotesize{}Savings} & \noun{\footnotesize{}Less}\tabularnewline
\end{tabular}\tabularnewline
\hline 
{\footnotesize{}Funders} & {\footnotesize{}}%
\begin{tabular}{c}
{\footnotesize{}Demand publications}\tabularnewline
\end{tabular} & {\footnotesize{}}%
\begin{tabular}{cc}
{\footnotesize{}Quality results} & \noun{\footnotesize{}Less}\tabularnewline
\end{tabular}\tabularnewline
\hline 
{\footnotesize{}Editors} & {\footnotesize{}}%
\begin{tabular}{c}
{\footnotesize{}Grant TA}\tabularnewline
\end{tabular} & {\footnotesize{}}%
\begin{tabular}{cc}
{\footnotesize{}Income} & \noun{\footnotesize{}More}\tabularnewline
\end{tabular}\tabularnewline
\hline 
{\footnotesize{}Politicians} & {\footnotesize{}}%
\begin{tabular}{c}
{\footnotesize{}Permit TA}\tabularnewline
\end{tabular} & {\footnotesize{}}%
\begin{tabular}{cc}
{\footnotesize{}Societal impact \& relevance} & \noun{\footnotesize{}Less}\tabularnewline
\end{tabular}\tabularnewline
\hline 
\end{tabular}{\footnotesize\par}
\par\end{center}

\begin{center}
Table 3: Interesting case 1, the pure current model
\par\end{center}

\begin{center}
{\footnotesize{}}%
\begin{tabular}{|c|c|c|}
\hline 
{\footnotesize{}Agents} & {\footnotesize{}Actions} & {\footnotesize{}Outcomes}\tabularnewline
\hline 
\hline 
{\footnotesize{}Academics} & {\footnotesize{}}%
\begin{tabular}{c}
{\footnotesize{}Publish OA}\tabularnewline
\end{tabular} & {\footnotesize{}}%
\begin{tabular}{cc}
{\footnotesize{}Opportunity} & \noun{\footnotesize{}Maximal}\tabularnewline
{\footnotesize{}Visibility} & \noun{\footnotesize{}More}\tabularnewline
{\footnotesize{}Prestige} & \noun{\footnotesize{}More}\tabularnewline
{\footnotesize{}Promotion} & \noun{\footnotesize{}More}\tabularnewline
\end{tabular}\tabularnewline
\hline 
{\footnotesize{}Administrators} & {\footnotesize{}}%
\begin{tabular}{c}
{\footnotesize{}Support OA}\tabularnewline
\end{tabular} & {\footnotesize{}}%
\begin{tabular}{cc}
{\footnotesize{}Savings} & \noun{\footnotesize{}More}\tabularnewline
\end{tabular}\tabularnewline
\hline 
{\footnotesize{}Funders} & {\footnotesize{}}%
\begin{tabular}{c}
{\footnotesize{}Demand OA publications}\tabularnewline
\end{tabular} & {\footnotesize{}}%
\begin{tabular}{cc}
{\footnotesize{}Quality results} & \noun{\footnotesize{}More}\tabularnewline
\end{tabular}\tabularnewline
\hline 
{\footnotesize{}Editors} & {\footnotesize{}}%
\begin{tabular}{c}
{\footnotesize{}Grant OA}\tabularnewline
\end{tabular} & {\footnotesize{}}%
\begin{tabular}{cc}
{\footnotesize{}Income} & \noun{\footnotesize{}Less}\tabularnewline
\end{tabular}\tabularnewline
\hline 
{\footnotesize{}Politicians} & {\footnotesize{}}%
\begin{tabular}{c}
{\footnotesize{}Demand green OA}\tabularnewline
\end{tabular} & {\footnotesize{}}%
\begin{tabular}{cc}
{\footnotesize{}Societal impact \& relevance} & \noun{\footnotesize{}More}\tabularnewline
\end{tabular}\tabularnewline
\hline 
\end{tabular}{\footnotesize\par}
\par\end{center}

\begin{center}
Table 4: Interesting case 2, the pure ideal model
\par\end{center}

Table 3 presents what we believe is characteristic of the current
situation. With overyone stuck to the TA option in their actions,
\uline{academics} are very limited in their abilities to learn
from contributions of others (as nobody can afford to buy them all)
and, therefore, their \uline{opportunity} to get published and
participate in the gran conversation is \uline{minimal}. Their
\uline{visibility} is, by the same reasoning, compromised (\uline{less}),
whereas for those who can actually get published (possibly by balancing
other interests not shown in this model) \uline{prestige} and \uline{promotion}
are guaranteed (\uline{more}). Toll access, TA, forces administrators
to pay more than they could for accessing collections. So, if they
actually \uline{support TA}, their institutional \uline{savings}
would be \uline{less}. \uline{Funders} who do request publications
as outcomes of the projects they fund would definitely have \uline{less}
\uline{quality results.} The only group clearly favored in this
setup is\uline{ editors }who sustain\uline{ more }income than
otherwise possible. Finally, by restricting opportunities to academics
and other members of the public to learn about contributions which
remain behind a paywall, politicians will see that the actual impact
and relevance of those contributions to society is lesser than possible. 

Table 4 paints a contrasting picture. \uline{Academic} commits
their publications to OA, empowering others to consult and use those
publications to support their own. Thus, the \uline{opportunity}
to publish is \uline{maximal}. But, of course, \uline{visibility},
\uline{prestige} and \uline{promotion} are favored with \uline{more}
opportunities for all. By encouraging and supporting OA, \uline{administrators}
contribute to \uline{more} budget \uline{savings} at their institutions.
\uline{Funders} will see \uline{more quality results} by \uline{demanding
OA publications}, not only because more publications will actually
be made, but also because more people, academics included, will have
opportunities to see and judge those results. This configuration,
of course, also determines that \uline{politicians} will see \uline{more}
\uline{impact and relevance} of results for society. Again, the
dissonats are the \uline{editors} who would more likely see \uline{less}
net \uline{income} in their regular bussiness. 

We must, of course, swiftly admit that we just made another simplifying
exercise very unsual in game theory. Instead of jumping to numerical
estimates of utilities, which could then be used to balance expected
utilities for the agents' strategies, by resorting to discrete domains
for the variables, we are doing a more qualitative analysis which
could be enlighting and does not rules out a traditional equilibrium
study. The resulting search space, however, it is still huge. There
are $110592=(3^{3}*4^{2}*2^{8})$ possible combinations of those variables-values.
But not all the combinations are meaningful in reality. 

We introduced some meaningful connections between the variables in
this model by devising a set of rules and constraint among their values
and running a tailored made contraint logic program on them. By these
means, we reduced the set to 3136 combinations. Some (26) of them
are shown in figure~\ref{combinationsofactions}.

\begin{center}
\begin{figure}
\begin{centering}
\label{combinationsofactions}
\includegraphics[scale=.4]{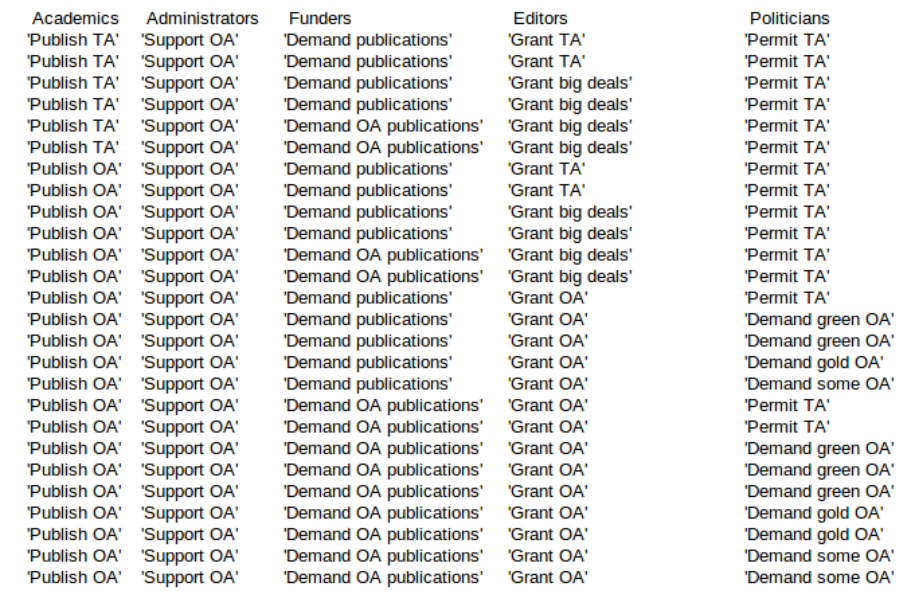}
\par\end{centering}
\caption{Some combinations of actions}
\end{figure}
\par\end{center}

This set of 26 action profiles is special for a number of reasons.
But before explaining, let us show the rules used to generate the
whole set: 

\begin{doublespace}
\textit{\noun{if Academics=`Publish TA' and Editors=`Grant TA' then
Academics' Opportunity = `Less' and Visibility =`Less'}}

\textit{\noun{if Academics=`Publish OA' and Editors=`Grant OA' then
Academics' Opportunity = `More' and Visibility =`More'}}

\textit{\noun{if Administrators =`Support OA' then Savings=`More',
otherwise Savings=`Less'. }}

\textit{\noun{if Funder=`Demand publications', Editors=`Grant TA'
and Politicians=`Permit TA' then Editor's Income = `More'.}}

\textit{\noun{if Editors=`Grant OA' then Editor's Income = `Less'.}}

\textit{\noun{if Editors=`Grant big deals' and Politicians=`Permit
TA' then Editor's Income = `More'.}}

\textit{\noun{if Funder=`Demand publications', Editors=`Grant TA'
and Politicians=`Permit TA' then Editor's Income = `More'.}}

\textit{\noun{if Editors=`Grant OA with embargoes' then Editor's Income
= `Less'.}}

\textit{\noun{if Funder=`Demand OA publications' then Editor's Income
= `Less'.}}

\textit{\noun{if Politicians=`Demand green OA' then Editor's Income
= `Less'.}}

\textit{\noun{if Visibility=`More' then Quality Results=`More' and
Impact and Relevance=`More'}}
\end{doublespace}

Over the thus reduced set of possibility, we move to consider the
regular next step in game theory modelling. Utility functions are
required for each agent to be able to compared strategy profiles and
search of equilibrium conditions. Once again, we try a very simple
approach taking advantage of the very simplified domains we have assigned
to the values of variables used to describe game's outcomes. Let us
assume that `More' values correspond to one (1) and `Less' values
to zero (0). A global utility function for this system could be:

\textit{GU = Opportunity + Visibility + Prestige + Promotion + Savings
+ Results + Income + Impact and Relevance}

A global utility function, however, is disregarded by the foundational
assumptions in game theory, which state that agents in a game could
not agree on a common, global set of preferences (and therefore utilities)
and must be assigned independent criteria for each. Otherwise, the
game would reduce to a standard optimization problem. 

Before we complain, let us point out that this global utility function
is, nevertheless, meaningful and it could indicate in our case, a
situation in which all the agent reach their higher benefits ($GU=8$,
in our simplified model). In fact, the 26 combinations in figure--\ref{combinationsofactions}
are the only ones that correspond to a GU=7, the highest value observed
in the whole set of 3136 pure strategic profiles. This fact is important,
as it makes us wonder whether all the agents in this game can be satisfied
and reach their highest levels of benefits. To answer this, we need
their particular utility functions.

Let us follow the simplified approach suggested by tables 2 to 4 and
define the corresponding utilities like this:

\textit{U}\textsubscript{Academics}\textit{ = Opportunity + Visibility
+ Prestige + Promotion}

\textit{U}\textsubscript{Administrators}\textit{ = Savings}

\textit{U}\textsubscript{Funders}\textit{ = Results}

\textit{U}\textsubscript{Editors}\textit{ = Income}

\textit{U}\textsubscript{Politicians}\textit{ = Impact and Relevance}

\begin{figure}
\begin{centering}
\includegraphics[scale=0.3]{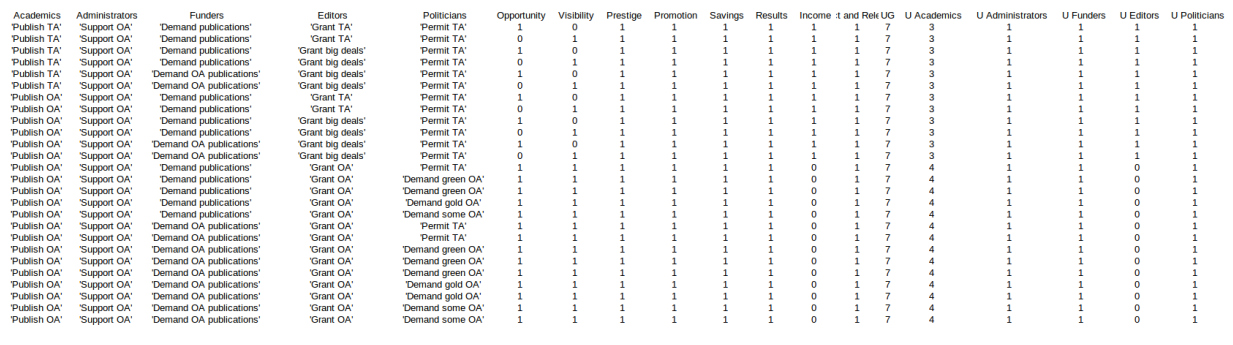}
\caption{Actions profiles with values and utilities}
\par\end{centering}
\label{actionsincomesutilities}
\end{figure}

Figure--\ref{actionsincomesutilities} shows the actions profiles
in the previous figure--\ref{combinationsofactions}together with
their outcome's values and utilities.

\subsection{Equilibrium conditions}

Are there conditions for equilibrium in this game?. Let us consider
the more common approach to equilibrium in game theory: Nash Equilibrium,
starting from the definition of best response by a player (From (Leyton-Shoham,
2012)):

\ovalbox{\begin{minipage}[t]{0.85\columnwidth}%
(Best Response) Player i's best response to the strategy profile $s_{-i}$
is a mixed strategy $s_{i}^{*}\in S_{i}$ such that $u(s_{i}^{*},s_{-i})\geq u(s_{i},s_{-i})$
for all strategies $s_{i}\in S_{i}$%
\end{minipage}}

In this context, each agent, a player, must be analized separately
and by means of strategy profiles which, in their purest form, are
the action combinations shown above (an pure profile is precisely
a row in figure--\ref{combinationsofactions}). The set of all strategy
profiles (which includes all the possible action combinations, but
also mixed strategies as explained above) is $S_{i}$. The variable
$s_{i}$ refers to the actions (possibly mixed strategy) of the player
under scrutiny and $s_{-i}$ are the actions (or strategies) of the
other agents in the game. Thus, a best response for agent i is an
action (or strategy) that produces the greatest utility (for agent
i, of course) among all its actions (or strategies) given the same
set of actions (or strategies) for the rest of the agents. 

We are now prepared for the definition of Nash Equilibrium:

\ovalbox{\begin{minipage}[t]{0.85\columnwidth}%
(Nash equilibrium) A strategy profile $s=(s_{1},\ldots,s_{n})$ is
a Nash equilibrium if for all agents i, $s_{i}$ is a best response
to $s_{-i}$%
\end{minipage}}

The reader must bear in mind that all these cumbersome references
to strategies instead of actions is due to the fact that the former
also include those mixed linear combinations of actions. 

Let us now try to analyze our game seeking Nash equilibria. As noticed
before, one of the practical difficulties in this game is that we
are dealing with a big set of agents and actions. However, one can
take further step of reduction and transform the game into many two-players
games, each of which would look like table 5. 

\begin{center}
\begin{tabular}{c|c|c|c|c|c|}
\multicolumn{1}{c}{} & \multicolumn{1}{c}{} & \multicolumn{1}{c}{} & \multicolumn{1}{c}{} & \multicolumn{1}{c}{Editors} & \multicolumn{1}{c}{}\tabularnewline
\cline{2-6} \cline{3-6} \cline{4-6} \cline{5-6} \cline{6-6} 
 &  & {\scriptsize{}big deals} & {\scriptsize{}TA} & {\scriptsize{}OA} & {\scriptsize{}OA with embargoes}\tabularnewline
\cline{2-6} \cline{3-6} \cline{4-6} \cline{5-6} \cline{6-6} 
Academics & {\scriptsize{}Publish TA} & \textbf{(3,1)} & \textbf{(3,1)} & \textit{(3,0)} & \textit{(3,0)}\tabularnewline
\cline{2-6} \cline{3-6} \cline{4-6} \cline{5-6} \cline{6-6} 
 & {\scriptsize{}Publish OA} & \textbf{(3,1)} & \textbf{(3,1)} & \textbf{(4,0)} & \textit{(3,0)}\tabularnewline
\cline{2-6} \cline{3-6} \cline{4-6} \cline{5-6} \cline{6-6} 
\end{tabular}
\par\end{center}

\begin{center}
Table 5
\par\end{center}

The outcomes values in table 5 has been obtained from the actual utility
computations of the original game. The ones shown in bold correspond
to profiles in which the global utility, as we defined above, is 7.
Whereas the ones in italic corresponds to profiles where that utility
is 6, as there no such kind of combination reaching 7. 

To make things clearer, let try one last simplification transforming
table 5 to an even simpler version:

\begin{center}
\begin{tabular}{c|c|c|c|c|}
\multicolumn{1}{c}{} & \multicolumn{1}{c}{} & \multicolumn{1}{c}{} & \multicolumn{1}{c}{Editors} & \multicolumn{1}{c}{}\tabularnewline
\cline{2-5} \cline{3-5} \cline{4-5} \cline{5-5} 
 &  & TA & OA & $EU_{Academics}$\tabularnewline
\cline{2-5} \cline{3-5} \cline{4-5} \cline{5-5} 
Academics & TA & \textbf{(3,1)} & \textit{(3,0)} & $3q$\tabularnewline
\cline{2-5} \cline{3-5} \cline{4-5} \cline{5-5} 
 & OA & \textbf{\{3,1\}} & \textbf{(4,0)} & $3q+4(1-q)$\tabularnewline
\cline{2-5} \cline{3-5} \cline{4-5} \cline{5-5} 
 & $EU_{Editors}$ & $p$ & $0$ & \tabularnewline
\cline{2-5} \cline{3-5} \cline{4-5} \cline{5-5} 
\end{tabular}
\par\end{center}

\begin{center}
Table 6
\par\end{center}

In table 6, we use \textit{q} to represent the probabity that Editors
play for TA action, that is one of Grant big deals or Grant TA, leaving
(1-\textit{q}) to indicate a OA action (Grant OA or Grant OA with
embargoes). We could have use \textit{p} to model the probabilities
for Academics, but, as the reader can verify, it is no necessary.

In this game, the strategy profile (\noun{Academics = OA, Editors
=TA}) is a Nash equilibrium. It is made of the best responses from
each agent to the other. 

\section{Conclusions}

Many conclusions can be drawn from the game here described. We have
made many simplyfing decisions along the modelling process that must
be taken into account as a determining context for those conclusions,
of course. It can be safely state, however, that under the current
conditions of the game, Editors will keep betting on Toll Access,
knowledge distribution models even if the whole set of Academics goes
for Open Access. 

\bibliographystyle{plainnat}
\phantomsection\addcontentsline{toc}{section}{\refname}\bibliography{aafloss}

\end{document}